\documentclass[3p,times]{elsarticle}

\usepackage{ecrc}
\usepackage{mathtools}
\usepackage[pdftex, colorlinks]{hyperref}

\volume{00}

\firstpage{1}

\journalname{Nuclear Physics A}

\runauth{K. Tywoniuk}

\jid{nupha}

\usepackage{amssymb}

\biboptions{sort&compress,numbers,comma,square}
 
\usepackage[figuresright]{rotating}

\def\bs{\boldsymbol}

\def\bdel{\boldsymbol \partial}

\def\q{{\boldsymbol q}}

\def\k{{\boldsymbol k}}

\def\n{{\boldsymbol n}}
\def\x{{\boldsymbol x}}
\def\y{{\boldsymbol y}}

\def\r{{\boldsymbol r}}
\def\z{{\boldsymbol z}}

\def\bkappa{{\boldsymbol \kappa}}

\def\tform{t_\text{f}}
\def\tdip{t_\text{d}}

\def\kform{k_\text{f}}

\def\Qmed{Q_s}

\def\R{\mathcal{R}}

\def\J{\mathcal{J}}

\def\P{\mathcal{P}}

\newcommand{\beq}{\begin{eqnarray}}
\newcommand{\eeq}{\end{eqnarray}}
\newcommand{\be}{\begin{eqnarray*}}
\newcommand{\ee}{\end{eqnarray*}}

\newcommand{\nn}{\nonumber\\ }

\begin{document}

\begin{frontmatter}



\dochead{}

\title{Advancing QCD-based calculations of energy loss}


\author{Konrad Tywoniuk}
\ead{konrad.tywoniuk@thep.lu.se}
\address{Department of Astronomy and Theoretical Physics,
Lund University,
S\"olvegatan 14A, 
SE-22 362 Lund, Sweden}

\begin{abstract}
We give a brief overview of the basics and current developments of QCD-based calculations of radiative processes in medium. We put an emphasis on the underlying physics concepts and discuss the theoretical uncertainties inherently associated with the fundamental parameters to be extracted from data. An important area of development is the study of the single-gluon emission in medium. Moreover, establishing the correct physical picture of multi-gluon emissions is imperative for comparison with data. We will report on progress made in both directions and discuss perspectives for the future. 
\end{abstract}


\end{frontmatter}

\section{Introduction}
\label{sec:intro}
The role of hard probes in heavy-ion collisions is twofold. Firstly, their creation involves short timescales, $\propto p_\perp^{-1}$ ($p_\perp$ being the transverse momentum), where plasma effects are negligible and, in case of electromagnetic probes which do not suffer further interactions, they can therefore serve as baselines for quantifying medium effects. For QCD probes, the second reason is that they are expected to hadronize at very large distances, $\propto p_\perp$. For analogous processes in the vacuum, this leaves a large window where the fragmentation process can be described purely in terms of perturbative QCD (pQCD), see e.g. \cite{Dokshitzer:1991wu}. The situation becomes much more complex in the presence of a medium, it being cold nuclear matter or a hot and dense plasma created in the aftermath of a heavy-ion collision. It involves additional screening and rescattering effects which, as experiments clearly show, modify the final-state hadron production quite significantly. Throughout, due to the strong separation of a typical large jet energy and a typical medium scale we will assume that these effects can be modeled within pQCD as well.

The purpose of this brief overview is not to provide a fully comprehensive account of the present status of the theory and phenomenology of jet quenching -- for that we refer the reader to recent reviews \cite{CasalderreySolana:2007zz,Majumder:2010qh} and references therein -- but rather to elucidate the basic concepts underlying our understanding of radiative processes in the medium. Along the way, we report on attempts toward higher precision or improved rigor in the understanding of several aspects of the process.  Attempts to quantify the differences between specific model implementations, which reflect a certain level of theoretical uncertainty, were already presented \cite{Armesto:2011ht}. Although most of the implementations agree on a qualitative level, the uncertainties presently complicate any quantitative statements about medium properties using hard probes. It is nevertheless encouraging to note the progress made along several directions which was reported at this conference.

At the outset, let us clarify the ``canonical" assumptions on the kinematics involved in the scattering and emission processes which, for our purpose, will set the stage for further discussion. Consider a projectile with energy $E$ emitting a secondary gluon with energy and transverse momentum $\omega$ and $k_\perp$, respectively. Interactions with the medium are purely transverse, given by $q_\perp$. Usually one assumes the following hierarchy of scales
\beq
\label{eq:kinematics-hierarchy}
E \gg \omega \gg k_\perp \sim q_\perp \,.
\eeq
Starting from the left, the two first inequalities are the so-called eikonal (soft) and collinear approximations, respectively, rendering the emission a purely semi-classical phenomenon. As will be discussed below, relaxing these in a well-controlled manner is a serious theoretical challenge but, at least for the former of the two, interpolating formulas are well-known. Going beyond the collinear approximation, on the other hand, is more delicate, see Sec.~\ref{sec:improvements} for further discussion. Finally, attempts to describe a more realistic medium interaction are also maturing, see Sec.~\ref{sec:thermal}.

\section{The single-gluon inclusive spectrum in medium}
\label{sec:onegluon}
Consider the propagation of a nascent, highly energetic parton through the medium. In this case, we have to treat bremsstrahlung and medium-induced gluon emissions on equal footing. Both of these components are contained in the rate\footnote{For a closely related approach, see also \cite{Guo:2000nz,Wang:2001ifa,Majumder:2009ge}.}
\begin{align}
\label{eq:independentspec-main}
\R &= 2\text{Re}  \int_0^\infty dt'\int_0^{t'} dt \int d^2\z \,\exp\left[-i \k \cdot \z-\frac{1}{2}\int_{t'}^\infty d\xi\, n(\xi) \sigma(\z) \right] \left. \bdel_y \cdot \bdel_z\,{\cal K}(t',\z\,;\,t, \y\,|\omega)\right|_{\y=0} \,,
\end{align}
where $n(t)$ is the medium density and
\beq
\label{eq:k-path-integral}
\mathcal{K}\left(t',\z; t,\y| \omega \right) = \int\mathcal{D} [{\bs r}] \, \exp\left\{ \int_{t}^{t'} \!\!d\xi \left[ i\frac{\omega}{2} \dot{{\bs r}}^2(\xi) - \frac{1}{2} n(\xi) \sigma({\bs r}) \right] \right\} \,,
\eeq
which is obtained by resumming multiple interactions with the medium \cite{Baier:1996kr,Baier:1996sk,Baier:1998kq,Zakharov:1996fv,Zakharov:1997uu,Wiedemann:2000za}.\footnote{A regularization prescription is needed to obtain the correct bremsstrahlung contribution \cite{Wiedemann:2000za}.} The integration variable ${\bs r}$ in \eqref{eq:k-path-integral} can be interpreted as the transverse size of the projectile-gluon fluctuation. The rate \eqref{eq:independentspec-main} is related to the spectrum of emitted gluons, see first term in \eqref{eq:antenna-spec1}, where $C_R$ is the appropriate color factor.\footnote{$C_R = C_F \equiv (N_c^2-1)/(2N_c)$ for the emission off a quark and $C_R = C_A \equiv N_c$ off a gluon, where $N_c = 3$ is the number of colors.} Due to the assumed strong separation of scales \eqref{eq:kinematics-hierarchy}, this spectrum does not account for the energy degradation of the particles along their trajectories. The medium-interactions are encoded in  the (dimensionless) cross section
\beq
\label{eq:DipoleMedium-CrossSection}
\sigma (\r) = \int \frac{d^2\q}{(2\pi)^2} \mathcal{V}^2(\q) \big[1 - \cos( \r \cdot \q)  \big] \,,
\eeq
where $\mathcal{V}^2(\q)$ is closely related to the elastic scattering rate, see Sec.~\ref{sec:thermal} for details. The spectrum can be solved analytically\footnote{For a numerical solution, interpolating between the two approximative schemes, see \cite{CaronHuot:2010bp}.} by expanding $\R$ up to fixed ($N$th) order in medium opacity $n(t) \sigma(\r)$ -- the so-called `opacity expansion' \cite{Gyulassy:2000fs,Gyulassy:2000er,Wiedemann:2000za} -- or by noting that for small arguments of the cosine in \eqref{eq:DipoleMedium-CrossSection} we can expand and write
\beq
\label{eq:harmonic-oscillator-approx}
n(t) \sigma(\r) \approx \frac{1}{2} \hat q(t) \r^2 \,,
\eeq
The latter procedure is referred to as the `harmonic oscillator' approximation and is only strictly valid for multiple soft scatterings, breaking down when the projectile becomes sensitive to isolated scattering centers. The transport coefficient $\hat q$, appearing in \eqref{eq:harmonic-oscillator-approx}, is further discussed in Sec.~\ref{sec:thermal}.

Before considering the medium-induced radiation, let us comment on the virtuality-driven bremsstrahlung component. In the absence of a medium, it is simply given by
\beq
\label{eq:r-brems}
\R\big|_\text{brems} = \frac{4 \omega^2}{\k^2} \,,
\eeq
and, as a rule, subtracted from the total spectrum to obtain a purely medium-induced quantity. In the presence of a medium, however, this contribution is in reality built up from both early- and late-time emissions \cite{MehtarTani:2012cy}. Thus, a careful treatment of this component is imperative for understanding the jets produced in heavy-ion collisions at LHC energies, see also \cite{Muller:2012hr} in these proceedings.

\subsection{Decoherence of radiation via the LPM mechanism}
\label{sec:decoherence}
The typical dipole size that governs the cross section in the exponent of \eqref{eq:DipoleMedium-CrossSection} is given by the characteristic size of the transverse medium fluctuation, which again is related to the screening length of the medium $\mu^{-1}$.\footnote{Inversely, $\mu$ denotes a typical transverse momentum exchanged with the medium.} In a dense medium, when the formation time of the gluon is large compared to the screening length, $t_\text{f} \gg \mu^{-1}$, the typical transverse momentum accumulated via coherent interaction with the medium is $\kform^2 = \sqrt{\omega \hat q}$, see e.g. \cite{MehtarTani:2012cy}. The formation time of this induced gluon reads $\tform = \sqrt{\omega /\hat q}$, and differs from the formation time in vacuum, which goes like $\tform = 1/(\omega \theta^2)$ and is typically long for soft gluons at a fixed angle $\theta$. In contrast, soft gluons are induced quite rapidly in the medium. It follows that the typical gluon fluctuation acquires a size $|{\bs r}| \sim t_\text{f} \theta_\text{f} \sim k_\text{f}^{-1}$, where we have used that $k_\text{f} \sim \theta_\text{f} \omega$, as could be directly anticipated. In this way the relevant gluon fluctuation mirrors the local medium characteristics. It follows that the radiated gluon completely decoheres from the projectile at times $t > \tform$.

In a medium of longitudinal extension $L$ and constant $\hat q$,\footnote{Solutions for a wide class of expanding media are also analytically availble \cite{Baier:1998yf,Arnold:2008iy}, see also \cite{Salgado:2002cd,Salgado:2003gb} for a general scaling law.} the dominant contribution to \eqref{eq:independentspec-main} arises when $\tform \ll L$. In the `harmonic oscillator' approximation the purely medium-induced rate can then be written as \cite{MehtarTani:2012cy}
\beq
\label{eq:r-in-in-final-leading}
\R\big|_\text{med-induced} \approx 4\omega  \int_0^L dt'  \int\frac{d^2\k'}{(2\pi)^2}\P(\k-\k',L-t')\sin\left(\frac{\k'^2}{2 \kform^2}\right)e^{-\frac{\k'^2}{2 \kform^2}}\,.
\eeq
It describes two stages: initially a (quantum) emission process of a gluon with transverse momentum strongly peaked around $\kform$, cf. two latter components (see also \cite{CaronHuot:2010bp}), convoluted with the probability
\beq
\label{eq:p-broadening}
\mathcal{P}(\k, \xi) \equiv  \frac{4\pi}{\hat q \xi} \exp \left[- \frac{\k^2}{\hat q \xi} \right] \,,
\eeq
describing the subsequent classical transverse momentum broadening of the gluon along the remaining path-length. This factor introduces the largest momentum scale of the problem, which we denote by $\Qmed = \sqrt{\hat q L}$. After integrating out the transverse momentum in \eqref{eq:r-in-in-final-leading}, we find the leading contribution to the medium-induced gluon energy spectrum
\beq
\label{eq:bdmps-estimate}
\omega \frac{dN}{d\omega} \big|_\text{med-induced} \simeq \frac{\alpha_s C_R}{\pi} \frac{L}{\tform} = \frac{\alpha_s C_R}{\pi} \sqrt{ \frac{\hat q L^2}{\omega} } \,.
\eeq
Thus, gluons with short formation times, $\tform \ll L$, can be produced anywhere along the medium length. In the opposite case, $\tform \gg L$, the estimate of (\ref{eq:bdmps-estimate}) breaks down and the spectrum is strongly suppressed. This is the so-called Landau-Pomeranchuk-Migdal (LPM) suppression in QCD.

In contrast to bremsstrahlung in vacuum \eqref{eq:r-brems}, the induced spectrum \eqref{eq:r-in-in-final-leading} is collinear safe and favors the emission of relatively soft gluons at very large angles via the broadening mechanism. This aspect challenges the use of the collinear approximation throughout and is a source of significant theoretical uncertainty \cite{Armesto:2011ht}.

\subsection{Modeling the medium}
\label{sec:thermal}
In the setup described above the input from an underlying theory of the plasma only enters as a two-point function of the medium gauge field. In the mixed representation
\beq
\label{eq:MediumAverage}
\langle \mathcal{A}^a_\text{med}(t, {\bs q}) \mathcal{A}^{\ast b}_\text{med}(t',{\bs q}')\rangle = \delta^{ab} \, n(t) \, \delta(t -t')\, (2\pi)^2 \delta^{(2)}(\q-\q') \, {\cal V}^ 2(\q) \,,
\eeq
where, in two simple cases, the squared potential is given by
\beq
{\cal V}^ 2(\q) = \left\{ \begin{array}{cl} \frac{m_D^2}{\left({\bs q}^2 + m_D^2 \right)^2} & \text{`static' scattering centers} \,, \\
                                                             \frac{m_D^2}{{\bs q}^2\left({\bs q}^2 + m_D^2 \right)} & \text{`dynamical' scattering centers} \,.\end{array} \right.
\eeq
The first model describes scattering off randomly distributed static, Yukawa potentials screened by a Debye mass $m_D$. It is often referred to as the Gyulassy-Wang model of the medium \cite{Gyulassy:1993hr}. The second option describes an elastic interaction with  medium quasi-particles in the HTL approximation of thermal QCD \cite{Aurenche:2002pd} where the screening mass is given by $m_D \sim gT$, $g$ being the coupling and $T$ the temperature.\footnote{In a thermal medium one also needs to take into account several subtle effects, e.g. such as an in-medium gluon mass.} In both cases the integral in \eqref{eq:DipoleMedium-CrossSection} is logarithmically divergent in the UV and one needs to introduce a cut-off -- another source of theoretical uncertainty \cite{Armesto:2011ht}. This form naturally appears for the derivation of \eqref{eq:independentspec-main} in thermal QCD \cite{Arnold:2001ba,Arnold:2001ms,Arnold:2002ja} and has also explicitly been calculated for the $N=1$ opacity case in \cite{Djordjevic:2007at,Djordjevic:2009cr}. 
One has also included finite chemical potential in the calculation \cite{Gervais:2012wd}.
Going beyond the HTL approximation, the authors of \cite{Djordjevic:2011dd} have also included a non-zero magnetic screening mass which is treated as a free parameter in their approach. The authors of \cite{Bluhm:2011sw,Bluhm:2012kp} have studied the effects of introducing a heuristic long-distance timescale of absorption which would affect the induction of hard gluons. Finally, we mention that the correlator can also be found using strong-coupling techniques, see e.g. \cite{Liu:2006he}.

Generally, the medium characteristics are encoded in the parameter $\hat q$ which represents the average transverse momentum broadening per unit length. It is known to leading order in QCD, see e.g. \cite{Arnold:2008vd}, but is usually treated as a phenomenological parameter to be extracted from data, see e.g. \cite{Armesto:2009zi}. Large next-to-leading corrections to this parameter was found in \cite{CaronHuot:2008ni}. Another higher-order effect is to account for the recoil of real or virtual gluon emissions \cite{Wu:2011kc}. Recently, there has been attempts to calculate it on the lattice \cite{Majumder:2012sh}. Ultimately, projectiles with lower longitudinal energy also become sensitive to medium drag effects \cite{Qin:2012cz} which become an additional source of energy loss.

\subsection{Further improvements on kinematics}
\label{sec:improvements}
Above, the eikonal approximation, $E \gg \omega$, has been employed, see \eqref{eq:kinematics-hierarchy}. In general, going beyond this picture implies taking care of underlying complications arising essentially from the uncertainty relation, see e.g. \cite{MehtarTani:2012hp}. Neglecting these subtleties, an interpolating formulation for all gluon energies up to $E\sim \omega$ is well-known in the literature for the rate in \eqref{eq:independentspec-main}, see e.g. \cite{Zakharov:1997uu,Apolinario:2012vy}.

As mentioned before, the inherent tendency toward large-angle emissions, $k_\perp \gtrsim \omega$, of the medium-induced single-gluon spectrum seriously questions the assumed small-angle approximation used for its derivation \eqref{eq:kinematics-hierarchy}. This poses a serious problem which can be considered \cite{Ovanesyan:2011xy,Ovanesyan:2011kn} within the so-called Soft-Collinear Effective Theory (SCET) augmented with new transverse soft degrees of freedom, so-called Glauber gluons which mimic the plasma interactions. For more details, see \cite{Idilbi:2008vm,DEramo:2010ak} and references therein. So far, sizable corrections has been reported for hard gluon emissions \cite{Ovanesyan:2011kn}. One the side, we note that the form of \eqref{eq:r-in-in-final-leading} suggests a separate treatment of the emission and broadening processes, also argued for in \cite{CasalderreySolana:2011rq}. A thorough analysis of these aspects could hopefully aid in reducing theoretical uncertainties and shed light on the space-time picture of the cascade.

\section{Interferences in medium}
\label{sec:interferences}
\begin{figure}[t!]
\centering
\begin{tabular}{c c c  c}
\begin{minipage}{4.5cm} 
\includegraphics[width=\textwidth]{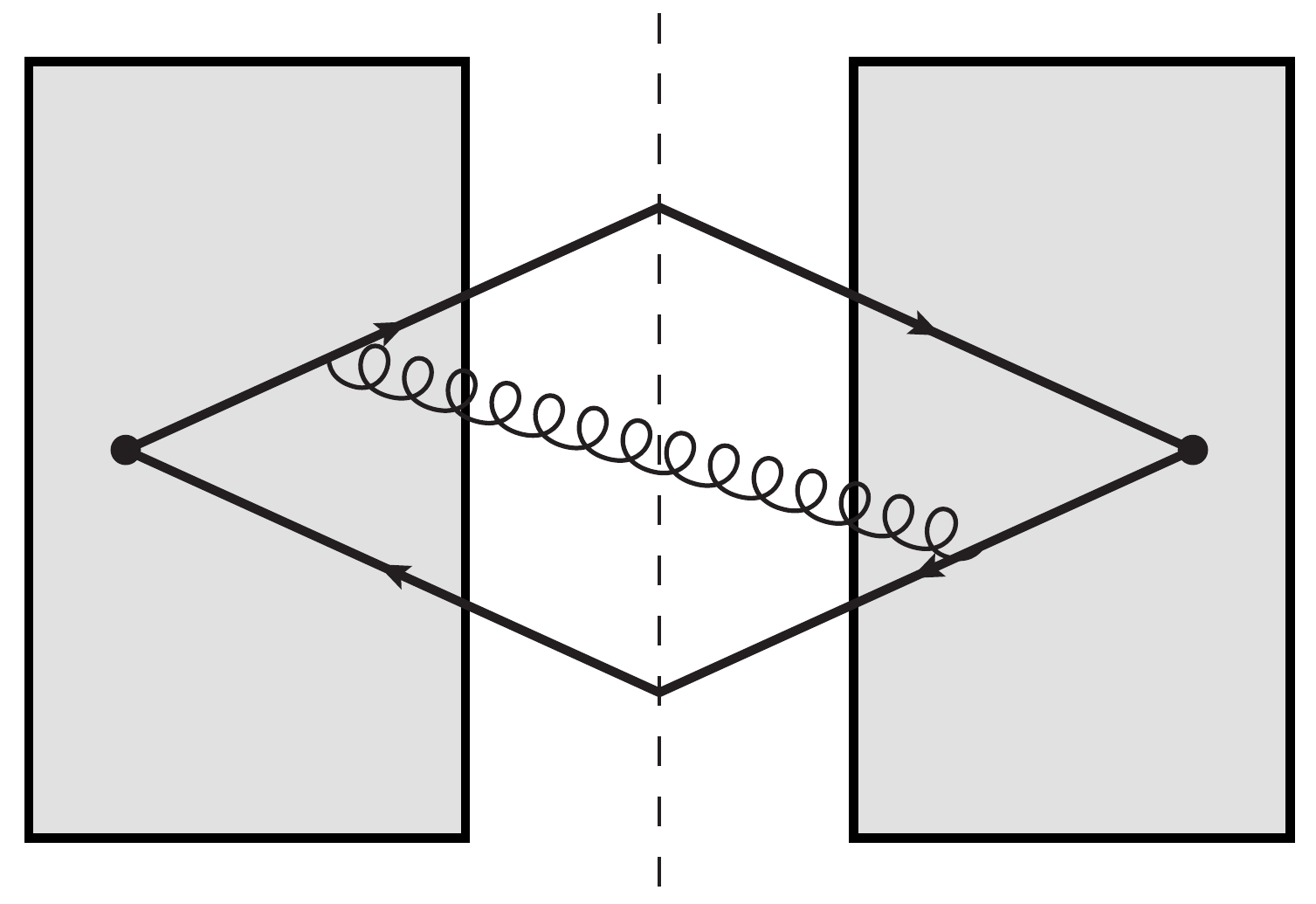}
 \end{minipage} &+&\begin{minipage}{4.5cm} 
\includegraphics[width=\textwidth]{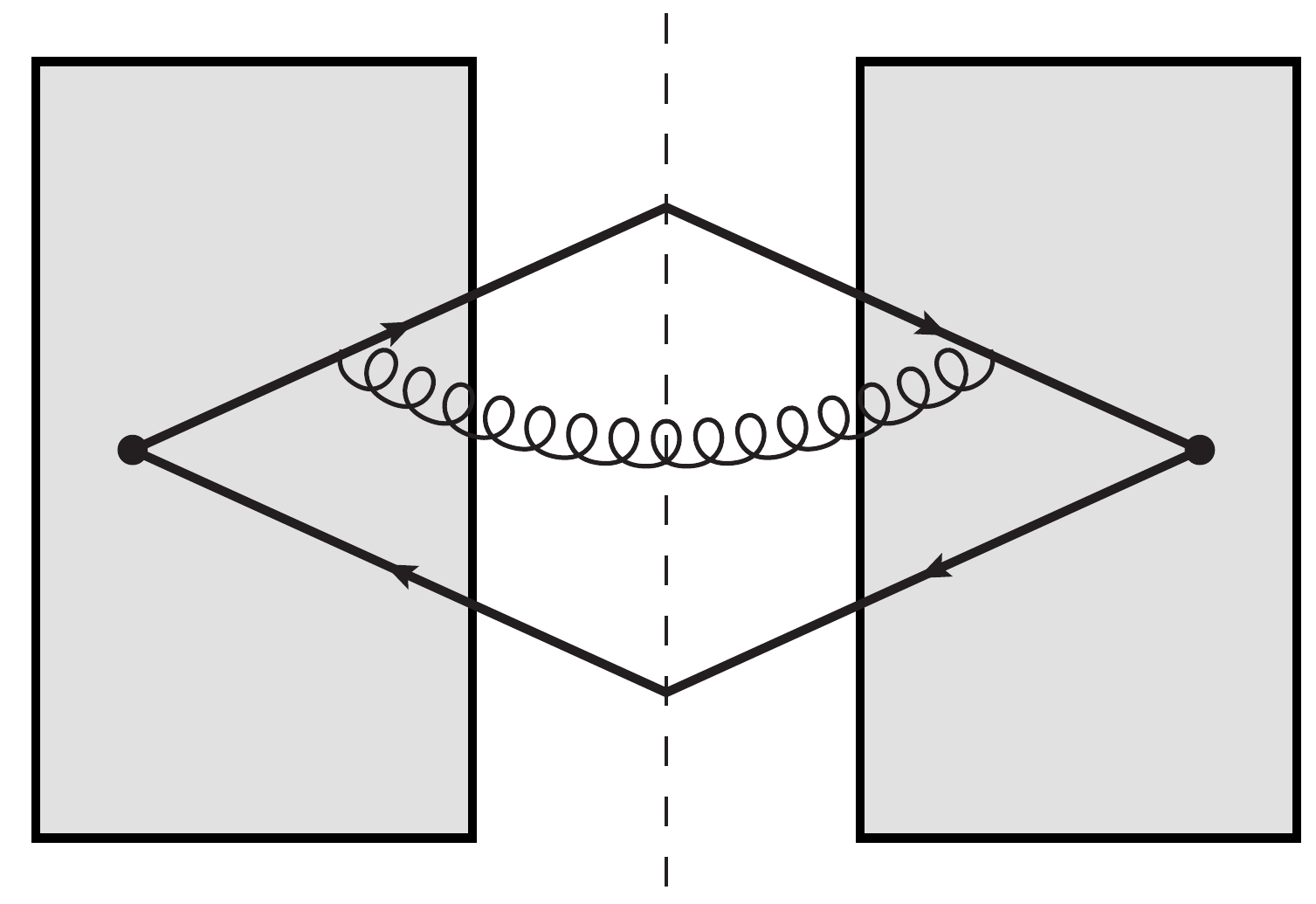}
 \end{minipage} &+sym.
\end{tabular}
\caption{Represention of the various contributions to the in-medium antenna spectrum. The first diagram stands for the interference part $\cal J$, while the second one stands for the independent spectrum of one of the legs, previously denoted ${\cal R}$. The rest symbolizes the symmetric configuration. }
\label{fig:CrossSection}
\end{figure}

So far we have only discussed the emission process taking place off a single charge propagating through the medium. However, at a given point during the jet cascading one would naturally expect the presence of many emitting charges, read jet fragments, which could give rise to interference effects. The simplest system which contains these interferences is the gluon radiation off two color charges ${\bs Q}_b$ and ${\bs Q}_c$, separated by an opening angle $\theta_{bc}$, originating from the decay of a virtual particle with color charge ${\bs Q}_a$.\footnote{E.g. for the process $g^\ast \to q \bar q \to q \bar q + g$, the squared charges are simply ${\bs Q}_a^2 = C_A$ and ${\bs Q}_b^2 = {\bs Q}_c^2 = C_F$.} The spectrum of the final gluon is called the `antenna spectrum' in the literature \cite{Dokshitzer:1991wu} and consists of direct emissions, described by the rate $\mathcal{R}$ as considered in the previous section, and of interferences, denoted by $\mathcal{J}$, see Fig.~\ref{fig:CrossSection}. The spectrum reads then
\begin{align}
\label{eq:antenna-spec1}
\omega \frac{dN}{d^3k} & = \frac{\alpha_s}{(2\pi)^2 \omega^2} \left[{\bs Q}_b^2 \mathcal{R}_b + {\bs Q}_c^2 \mathcal{R}_c + 2 {\bs Q}_b\cdot {\bs Q}_c \mathcal{J} \right] \,, \\
\label{eq:antenna-spec2}
& = \frac{\alpha_s}{(2\pi)^2 \omega^2} \left[{\bs Q}_a^2 \mathcal{J} + {\bs Q}_b^2 \left(\mathcal{R}_b - \mathcal{J} \right) +{\bs Q}_c^2 \left(\mathcal{R}_c - \mathcal{J} \right) \right] \,,
\end{align}
due to color conservation ${\bs Q}_a + {\bs Q}_b + {\bs Q}_c = 0$. The last two terms in \eqref{eq:antenna-spec2} are the so-called coherent spectra off charges $b$ and $c$, respectively. The first term, on the other hand, describes emissions off the total charge and is a manifestation of color conservation and coherence. Compared to our previous analysis, the novel term is contained in the interferences, which in the presence of a medium read \cite{MehtarTani:2011jw,CasalderreySolana:2011rz}
\begin{align}
\label{eq:interferencespec-main}
\J&= \text{Re}  \int_0^\infty dt'\int_0^{t'} dt \; \big[1-\Delta_{\text{med}}(t) \big] \int d^2\z \,\exp\left[-i \bkappa\cdot \z-\frac{1}{2}\int_{t'}^\infty d\xi\, n(\xi) \sigma(\z)+i\frac{\delta \k^2}{2 \omega} t\right] \nn
& \qquad \times \left.  \left(\bdel_y-i\delta \k\right)\cdot \bdel_z\,{\cal K}(t',\z\,;\,t, \y\,|\omega)\right|_{\y=\delta\n t} +\text{sym.} \,,
\end{align}
where $|\delta \k | = \omega |\delta \n| \sim \omega\theta_{bc}$ is related to the opening angle of the antenna.\footnote{The antenna spectrum has also been calculated for massive quarks in \cite{Armesto:2011ir}.} The main novel component is the so-called decoherence parameter, which in the `harmonic oscillator' approximation discussed above reads
\beq
\label{eq:decoh-parameter}
1-\Delta_\text{med}(t) = e^{- \hat q t \,\left(\theta_{bc} t \right)^2} \,,
\eeq
where $\theta_{bc} t$ is the transverse size of the antenna prior to emission. The argument of the exponent defines a characteristic timescale of decoherence of the antenna system $t_\text{d}$, given by $t_\text{d} = (\hat q \theta_{bc}^2)^{-1/3}$. It is equivalent to the formation time for induced gluons, $\tform$, in the sense that for $t \gg t_\text{d}$ the antenna legs decohere and behave as independent objects, i.e. $\mathcal{J} \to 0$.

\begin{figure}[t]
\centering
\includegraphics[width=0.35\textwidth]{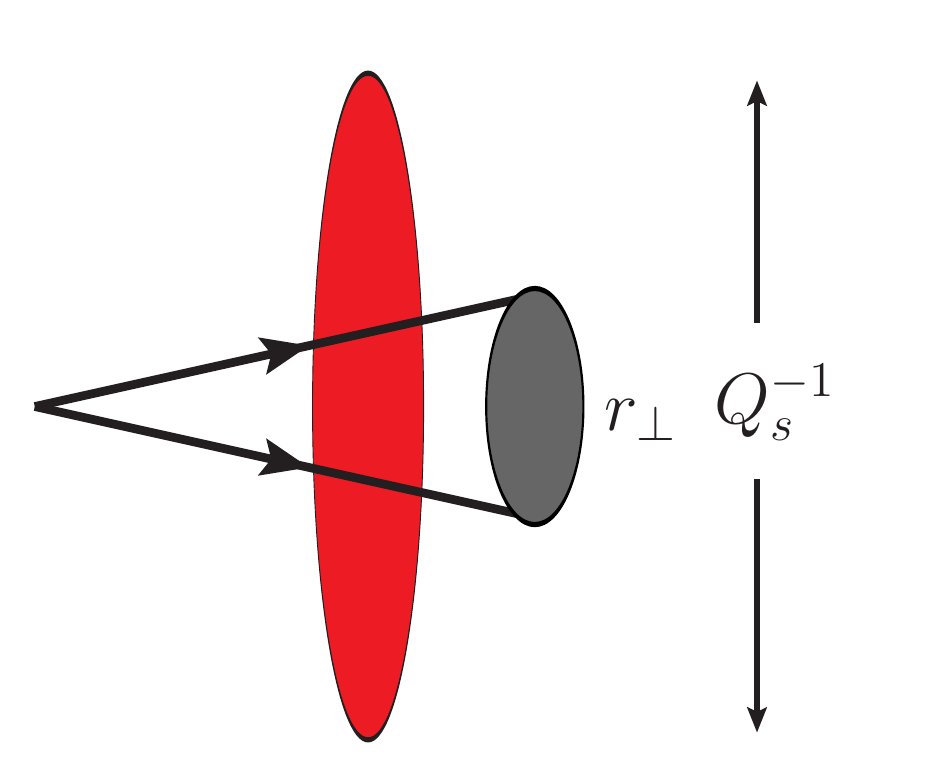}\hspace{1cm}
\includegraphics[width=0.28\textwidth]{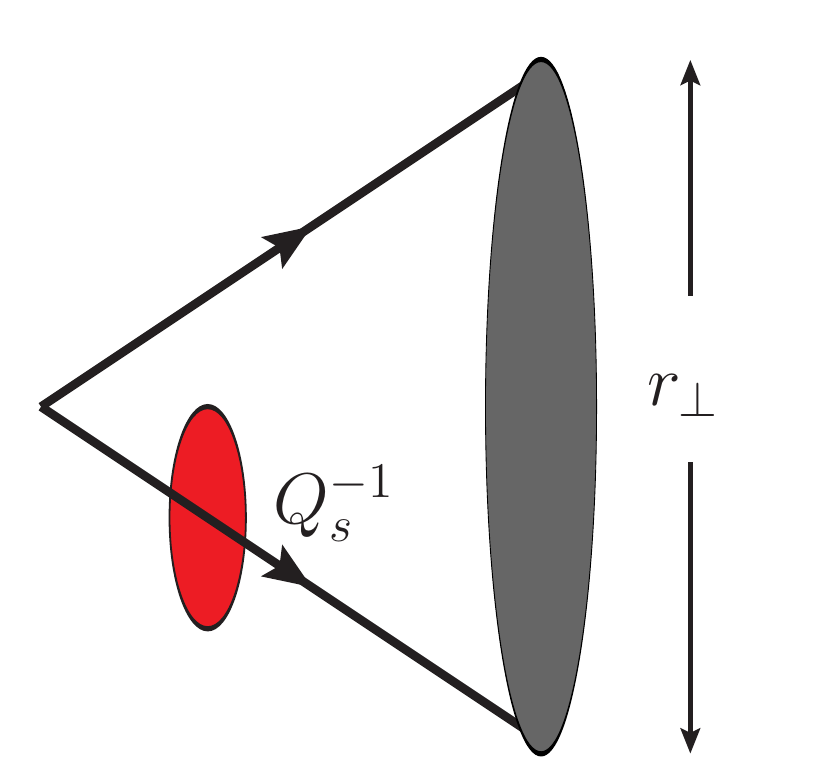}
\caption{The characteristic regimes of radiation in media: the ``dipole" regime, $r_\perp \ll Q_s^{-1}$ (left) and the ``saturated" regime, $r_\perp \gg Q_s^{-1}$ (right).}
\label{fig:regimes}
\end{figure}

The behavior of the system is governed by the hardest scale of the system, which is one of the following three characteristic scales \cite{MehtarTani:2011gf,MehtarTani:2012cy}
\beq
\label{eq:hard-scale}
Q_\text{hard} = \max \left(Q_s, r_\perp^{-1}, \omega|\delta \n| \right) \,,
\eeq
which determines the largest transverse momentum that can be generated by the system (here $r_\perp \sim \theta_{bc} L$ is the maximal antenna size). Thus, in the general case we expect the spectrum to be suppressed for $k_\perp > Q_\text{hard}$. Starting with the latter, this is the scale that governs angular ordering in vacuum, i.e. when the coherent spectrum is suppressed at angles $\theta > \theta_{bc}$, see also \cite{Beraudo:2012hs}.
The two former scales arise exclusively in the case of a medium (let us presently assume that both are much larger than $\omega |\delta \n|$). Their relationship defines two distinct regimes, the ``dipole" and ``decoherence" regimes, respectively, which can be recast geometrically, illustrated in Fig.~\ref{fig:regimes}.

In the ``dipole" regime the antenna remains in a coherent state, $\tdip \gg L$, and radiates accordingly, i.e. the medium-induced independent spectra are cancelled. A novel bremsstrahlung component, proportional to $\Delta_\text{med} \sim r_\perp^2$, arises between the opening angle and the maximal angle $(\omega r_\perp)^{-1}$ \cite{MehtarTani:2010ma,MehtarTani:2011gf,MehtarTani:2012cy}. In the ``decoherence" regime, on the other hand, the antenna constituents rapidly decohere and radiate almost completely independently, see also \cite{CasalderreySolana:2011rz}. Above the maximal angle, in this case $Q_s/\omega$, both the medium-induced and the bremsstrahlung components drop rapidly marking the onset of coherence. For more details, see \cite{MehtarTani:2011gf,MehtarTani:2012cy}.
Note that in the soft sector the maximal angle is arbitrary large. The behavior becomes universal and gives rise to a very simple coherent spectrum, namely \cite{MehtarTani:2010ma,MehtarTani:2011tz}
\beq
\label{eq:nqmed}
dN \big|_\text{soft, coh} = \frac{\alpha_s C_R}{\pi}\frac{d\omega}{\omega}\frac{\sin\theta \ d \theta}{1-\cos\theta}\Big[\Theta(\cos\theta-\cos\theta_{q\bar q}) 
+\Delta_{\text{med}}\,\Theta(\cos\theta_{q\bar q}-\cos\theta)\Big] \,.
\eeq
We observe that a novel, infrared divergent component, which is absent from the independent spectrum \eqref{eq:r-in-in-final-leading}, is found in the case of two emitters. Thus, for soft gluons in dense media, $\Delta_\text{med} \sim 1$, all charges radiate independently (also the radiation off the total charge cancels out) due to the strong screening of the medium \cite{MehtarTani:2011tz}. The aspects discussed here will hopefully allow for a more rigorous derivation of multi-gluon processes in the presence of a medium, which we proceed to discuss.

\section{Multi-gluon emissions}
\label{sec:multigluon}
In addition to the above-mentioned inherent uncertainties of the single-gluon spectrum, another serious source of uncertainty arises when comparing to high-$p_\perp$ hadron spectra in relativistic heavy-ion collisions or, even more so, jets and jet substructure, where effects of multiple radiation (especially in the latter cases) necessarily have to be accounted for. In lieu of a rigorously derived fragmentation function, or jet calculus, in medium one usually employs working models, which a few of the most common we briefly outline below. For a more thorough discussion, see also \cite{Majumder:2010qh,Armesto:2011ht}.

Firstly, one can assume that the medium effects factorize into a probability of energy loss to be convoluted with a vacuum fragmentation function \cite{Baier:2001yt,Salgado:2003gb}, see \cite[Sec. 1]{Beraudo:2012hs} in these proceedings. Such a convolution naturally favors the medium-induction of hard gluon emissions in \eqref{eq:r-in-in-final-leading}. As already discussed in \cite{Beraudo:2012hs} certain medium effects, e.g. modified color flow, could invalidate such a factorized scheme. Another approach in the literature has been to define a modified Altarelli-Parisi splitting function, see \cite{Polosa:2006hb} and also \cite{Guo:2000nz,Wang:2001ifa,Majumder:2009ge}, which includes the effects of medium-induced radiation into the DGLAP cascade.\footnote{These approaches are in fact closely related \cite{Armesto:2007dt}; by neglecting the virtuality ordering along the DGLAP cascade on comes back to the factorization lying at the heart of the quenching weights.} This procedure has been implemented in Monte-Carlo generators such as Q-PYTHIA \cite{Armesto:2009fj}, Q-HERWIG \cite{Armesto:2009ab} and others. Finally, a formulation of the medium-induced emissions in terms of a time-dependent rate equation \cite{Jeon:2003gi,CaronHuot:2010bp} has been implemented in the Monte-Carlo generator MARTINI \cite{Schenke:2009gb}. Worth mentioning here is also the versatile Monte-Carlo generator JEWEL \cite{Zapp:2008af,Zapp:2011ya} which aims at a microscopical modeling of both the medium interactions and the subsequent induced radiation and which incorporates the LPM effect probabilistically.

To date none of the above approaches attempt to treat possible interference effects, discussed in Sec.~\ref{sec:interferences}. Focussing on processes where these are naturally suppressed, i.e., for soft gluon production in dense media ($t_\text{f} \ll L$), one has recently formulated a generating functional for multi-gluon emissions, see \cite{MehtarTani:2012hp} in these proceedings. Besides, as mentioned before, the interface between bremsstrahlung and medium-induced radiation is unchartered and highly model dependent. We anticipate that progress along these lines can further elucidate the space-time structure of in-medium radiation and help putting calculations of jet cascading in heavy-ion collisions on firmer ground.

\section{Perspectives}
\label{sec:perspectives}
The calculation of radiative processes in medium has been established as a coherent framework within perturbative QCD. In this overview we have focussed on a few areas of progress. Firstly, improvements on the kinematics of emission which would be important for secondary emissions. Secondly, a more rigorous treatment of parton propagation in a QCD background which is imperative for a proper interpretation of the extracted medium characteristics from data. And, finally, attempts toward developing a in-medium multi-gluon calculus and analyses of interference effects. Hopefully, these improvements will elucidate our understanding of the dynamics of the plasma and aid in the construction of detailed models to be confronted with experimental data. \\

\emph{Acknowledgements:}
I would like to thank Y.~Mehtar-Tani and C.~A.~Salgado for collaboration and fruitful discussions, and N.~Armesto and N.~Su for helpful comments on the manuscript. This work was supported in part by the Swedish Research Council (contract number 621-2010-3326). 


\bibliographystyle{elsarticle-num}
\bibliography{hp2012}


\end{document}